\documentclass[fleqn,12pt]{wlscirep}
\usepackage{wrapfig, framed, caption}
\usepackage[utf8]{inputenc}
\usepackage{heuristica}
\usepackage{graphicx}
\usepackage{wrapfig, framed, caption}
\usepackage{tcolorbox}
\usepackage{xr}
\usepackage{hyperref}
\usepackage{xcolor}

\makeatletter
\newcommand*{\addFileDependency}[1]{
  \typeout{(#1)}
  \@addtofilelist{#1}
  \IfFileExists{#1}{}{\typeout{No file #1.}}
}
\makeatother

\usepackage{setspace}
\usepackage[normalem]{ulem}
\useunder{\uline}{\ulined}{}
\usepackage{xparse}
\newsavebox{\fminipagebox}
\NewDocumentEnvironment{fminipage}{m O{\fboxsep}}
 {\par\kern#2\noindent\begin{lrbox}{\fminipagebox}
  \begin{minipage}{#1}\ignorespaces
 \end{minipage}\end{lrbox}%
  \makebox[#1]{%
    \kern\dimexpr-\fboxsep-\fboxrule\relax
    \fbox{\usebox{\fminipagebox}}%
    \kern\dimexpr-\fboxsep-\fboxrule\relax
  }\par\kern#2
 }

\title{FIFA World Cup 2022 – The Network Edition}

\author[*1,2,3]{Mil\'an Janosov}
\author[4]{Patrik Szigeti}

\affil[1]{Department of Network and Data Science, Central European University, Budapest, Hungary}

\affil[2]{Baoba Inc., Delaware, United States}

\affil[3]{Milan Janosov \href{https://linktr.ee/janosov}{https://linktr.ee/janosov}}

\affil[4]{Revolut, Budapest, Hungary}

\affil[*]{janosovm@ceu.edu}

\begin{document}

\maketitle


\section*{Abstract}
{\small

After a long qualifying process packed with surprises (Italy missing out as the reigning European champions) and last minute drama (both Egypt and Peru missed out on penalties), the FIFA World Cup 2022 kicked off on the 20th of November in Qatar. With 32 countries and over 800 players representing nearly 300 clubs globally, it measured up to more than 12 billion EUR in the players’ current estimated market value total. In this short piece, we explore what the small and interconnected world of football stars looks like and even make a few efforts and compare success in soccer to social networks. 

}

\vspace{0.5cm}
{\small {\bf Keywords}: network science, social network analysis, soccer, data science}

\vspace{1.0cm}
{\it \hspace{-1cm} Published in Nightingale, Journal of the Data Visualization Society, December 23, 2022~\cite{nightingale}. }
\vspace{1.0cm}

\section{Data}

We are data scientists with a seasoned football expert on board, so we went for one of the most obvious choices of the field – www.transfermarkt.com. We first wrote a few lines of Python code to scrape the list of participating teams~\cite{teams}, the list of each team’s players~\cite{players}, and the detailed club-level transfer histories of these players arriving at the impressive stats of our intro by comprising the complete transfer history of 800 players, measuring up to 6,600 transfers and dating back to 1995 with the first events.

\section{Club network}

The majority of players came from the top five leagues (England, Spain, Italy, Germany, and France) and represented household teams such as Barcelona (with 17 players), Bayern Munich (16), or Manchester City (16). While that was no surprise, one of the many wonders of a World Cup is that players from all around the globe can show their talents. Though not as famous as the ‘big clubs’, Qatari Al Sadd gave 15 players, more than the likes of Real Madrid or Paris Saint-Germain! There are, however, great imbalances when throwing these players’ market values and transfer fees into the mix. To outline these, we decided to visualize the typical ‘migration’ path football players follow – what are the most likely career steps they make one after the other?

A good way to capture this, following the prestige analysis of art institutions~\cite{fraiberger2018quantifying}, is to introduce network science~\cite{netsci} and build a network of football clubs. In this network, every node corresponds to a club, while the network connections encode various relationships between them. These relationships may encode the interplay of different properties of clubs, where looking at the exchange of players (and cash) seems a natural choice. In other words, the directed transfers of players between clubs tie the clubs into a hidden network. Due to its directness, this network also encodes information about the typical pathways of players via the ‘from’ and ‘to’ directions, which eventually capture the different roles of clubs as attractors and sinks.

\begin{table}[!hbt]
\centering
  \includegraphics[scale=0.5]{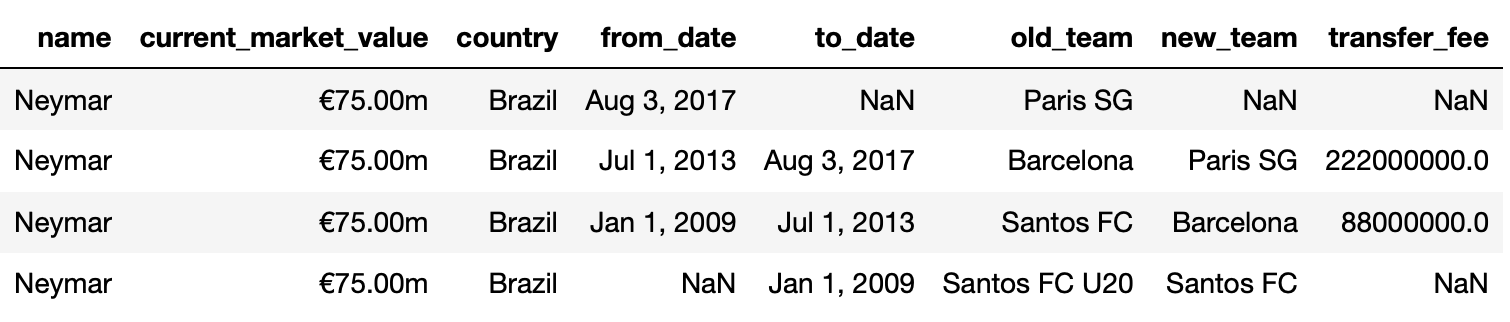}
  \caption{The datafied transfer history of Neymar.}
  \label{tab:tab1}
\end{table}

To do this in practice, our unit of measure is the individual transfer history of each player, shown in Table \ref{tab:tab1} for the famous Brazilian player known simply as Neymar. This table visualizes his career trajectory in a datafied format, attaching dates and market values to each occasion he changed teams. His career path looks clean from a data perspective, although football fans will remember that it was anything but – his fee of EUR 222M from Barcelona to PSG still holds the transfer record to this day. These career steps, quantified by the transfers, encode upgrades in the case of Neymar. In less fortunate situations, these prices can go down signaling a downgrade in a player’s career. 

\begin{figure}[!hbt]
\centering
\includegraphics[width=0.75\textwidth]{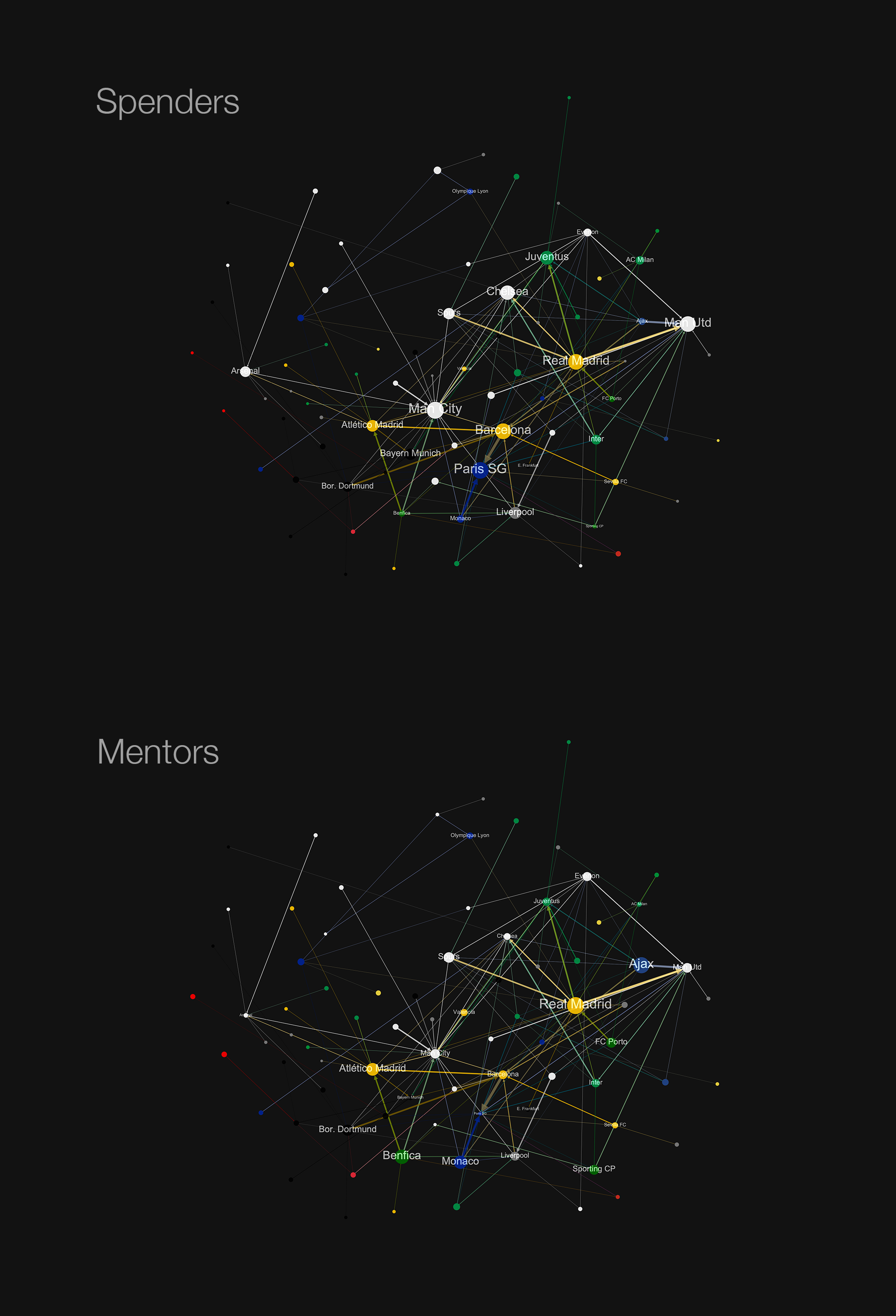}
\caption{The network of the top football clubs based on the total amount of money spent and received on player transfers. Node sizes correspond to these values, while node coloring shows the dominant color of each club’s home country flag.}
\label{fig:fig1}
\end{figure}

\clearpage

Following this logic in our analysis, we assumed that two clubs, A and B, were linked (the old and new teams of a player), if a player was transferred between them, and the strength of this link corresponded to the total amount of cash associated with that transaction. The more transactions the two clubs had, the stronger their direct connection was (which can go both ways), with a weight equal to the total sum of transfers (in each direction). In the case of Neymar, this definition resulted in a direct network link pointing from Barcelona to Paris SG with a total value of EUR 222M paid for the left winger.

Next, we processed the more than six thousand transfers of the 800+ players and arrived at the network of teams shown in Figure \ref{fig:fig1}. To design the final network, we went for the core of big money transactions and only kept network links that represented transfer deals worth more than EUR 2.5M in total. This network shows about 80 clubs and 160 migration channels of transfers. To accurately represent the two aspects of transfers (spending and earning) we created two versions of the same network. The first version measures node sizes as the total money invested in new players (dubbed as spenders), while the second version scales nodes as the total money acquired by selling players (dubbed as mentors).

\paragraph{Spenders.} 

The first network shows us which clubs spent the most on players competing in the World Cup, with the node sizes corresponding to the total money spent. You can see the usual suspects: PSG, the two clubs from Manchester, United, and City, and the Spanish giants, Barcelona, and Real Madrid. Following closely behind are Chelsea, Juventus, and Liverpool. It’s interesting to see Arsenal, who – under Arteta’s management – can finally spend on players, and Bayern Munich, who spend a lot of money but also make sure to snatch up free agents as much as possible.

Explore these relationships and the network in more detail by looking at Real Madrid! Los Blancos, as they’re called, have multiple strong connections. Their relationship with Tottenham is entirely down to two players who played an integral part in Real Madrid’s incredible 3-year winning spell in the Champions League between 2016 and 2018: Croatian Luka Modric cost 35M, and Welsh Gareth Bale cost an at-the-time record-breaking 101M. While Real Madrid paid 94M for Cristiano Ronaldo in 2009 to Man Utd, in recent years there was a turn in money flow, and United paid a combined 186M for three players: Ángel Di María, Raphael Varane, and Casemiro. They also managed to sell Cristiano Ronaldo with a profit to Juventus for 117M.

One can see other strong connections as well, such as Paris SG paying a fortune to Barcelona for Neymar and Monaco for Kylian Mbappé. There are also a few typical paths players take – Borussia Dortmund to Bayern Munich, Atlético Madrid to Barcelona, or vice versa. It’s also interesting to see how many different edges connect to these giants. Man City has been doing business worth over EUR 1M with 27 different clubs.

\paragraph{Mentors.} 

The second network shows which clubs grow talent instead of buying them and have received a substantial amount of money in return. Node sizes represent the amount of transfer fees received. This paints a very different picture from our first network except for one huge similarity: Real Madrid. In the past, they were considered the biggest spenders. They have since adopted a more business-focused strategy and managed to sell players for high fees as mentioned above.

A striking difference, however, is while the top spenders were all part of the top five leagues, the largest talent pools came from outside this cohort except for Monaco. Benfica, Sporting, and FC Porto from Portugal, and Ajax from the Netherlands are all famous for their young home-grown talents, and used as a stepping stone for players from other continents. Ajax has sold players who competed in this World Cup for over EUR 560M. Their highest received transfer fees include 85.5M for Matthijs de Ligt from Juventus and 86M for Frenkie de Jong from Barcelona. Ajax signed de Jong for a total of EUR 1 from Willem II in 2015 when he was 18, and de Light grew up in Ajax’s famous academy. Not to mention that they recently sold Brazilian Antony to Manchester United for a record fee of 95M. They paid 15.75M for him just 2 years ago – that’s almost 80M in profit. Insane!

Benfica earned close to 500M, most recently selling Uruguayan Darwin Nunez for 80M to Liverpool. The record fee they received is a staggering 127M for Portuguese Joao Félix from Atlético Madrid, who grew up at Benfica. Monaco earned 440M from selling players such as Kylian Mbappé (180M) and Aurélien Tchouameni (80M), Portuguese Bernardo Silva (50M), Brazilian Fabinho and Belgian Youri Tielemans (both for 45M). These clubs have become incredible talent pools for the bigger clubs, therefore really appealing to young players. It’s interesting to see how many edges the nodes for these clubs have, further proving that these teams function as a means for reaching that next level.

\section{Player network}

After looking at the club-to-club relationships, zoom in on the network of players binding these top clubs together. Here, we built on the players’ transfer histories again and reconstructed their career timelines. Then we compared these timelines between each pair of World Cup players, noted if they ever played for the same team, and if so, how many years of overlap they had (if any).

\begin{figure}[!hbt]
\centering
\includegraphics[width=0.90\textwidth]{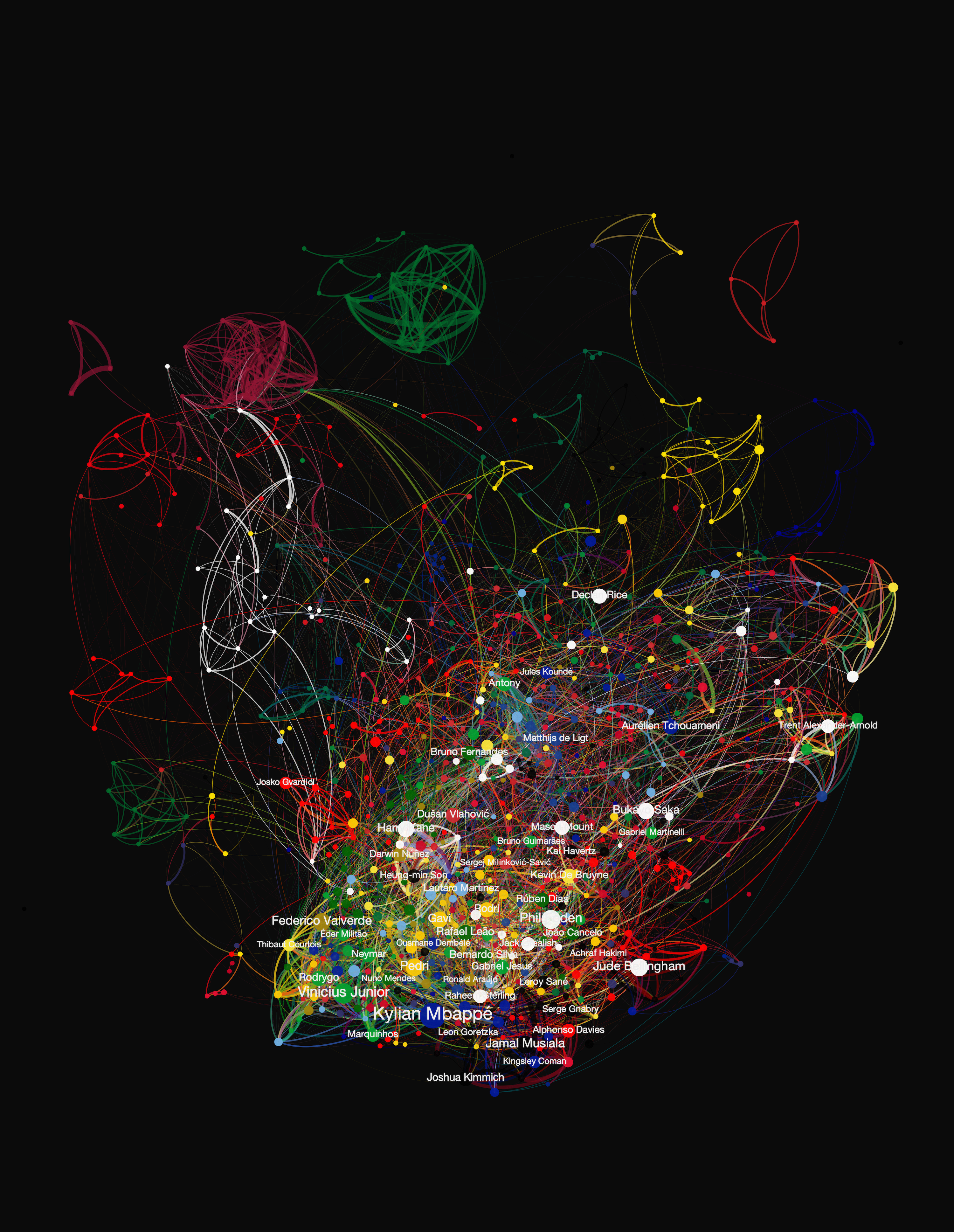}
\caption{The player-level network showing previous and current teammate relationships. Note size corresponds to the players’ current market values, while color encodes their nationality based on their country’s flag’s primary color. See the interactive version of this network here~\cite{interactive}.}
\label{fig:fig2}
\end{figure}

To our biggest surprise, we got a rather intertwined network of 830 players connected by about 6,400 former and current teammate relationships, as shown in Figure \ref{fig:fig2}. Additionally, the so-called average path length turned out to be 3 – which means if we pick two players at random, they most likely both have teammates who played together at some point. Node sizes were determined by a player’s current market value, and clusters were colored by the league’s nation where these players play. It didn’t come as a surprise that current teammates would be closer to each other in our network. You can see some interesting clusters here, with Real Madrid, Barcelona, PSG, and Bayern Munich dominating the lower part of the network and making up its center of gravity.

Why is that? The most valuable player of the World Cup was Kylian Mbappé, with a market value of 160M, surrounded by his PSG teammates like Brazilians Marquinhos and Neymar and Argentinian Lionel Messi. Messi played in Barcelona until 2021, with both Neymar and Ousmane Dembélé connecting the two clusters strongly. Kingsley Coman joined Bayern Munich in 2017, but he played for PSG up until 2014, where they were teammates with Marquinhos, thus connecting the two clusters.

You can discover more interesting patterns in this graph, such as how the majority of the most valuable players have played together directly or indirectly. You can also see Englishmen Trent Alexander-Arnold (Liverpool) or Declan Rice (West Ham United) further away from the others. Both of those players only ever played for their childhood clubs. But the tight interconnectedness of this network is also evident with how close Alexander-Arnold actually is to Kylian Mbappé. During the 2017–2018 season, Mbappé played at Monaco with Fabinho behind him in midfield, who signed for Liverpool at the end of the season, making him and Alexander-Arnold teammates.

\clearpage

With the World Cup hosting hundreds of teams’ players from various nations, there are obviously some clusters that won’t connect to these bigger groups. Many nations have players who have only played in their home league, such as this World Cup’s host nation Qatar (maroon cluster in the top left corner). Saudi Arabia (green cluster next to Qatar) beat Argentina, causing one of this year’s biggest surprises. Morocco (red cluster in the top right corner) delivered the best-ever performance by an African nation in the history of the World Cups. Both of those nations join Qatar in this category of home-grown talent. These players will only show connections if they play in the same team – in the case of the Moroccan cluster, that team is Wydad Casablanca. The Hungarian first league’s only representative at the World Cup, Tunisian Aissa Laidouni from Ferencváros hasn’t played with anyone else on a club level who has made it to the World Cup. He became a lone node on our network. That shouldn’t be the case for long, considering how well he played in the group stages.

\section{Success and networks}

The potential success of different teams and the outcomes of championships have been majorly interesting for data and statistics people since the era of Moneyball. Ever since, a wide range of efforts came to light about the possibilities to predict the outcomes of sporting events, from asking actual animals like Paul the Octopus to serious academic research, and even companies specializing in this domain.

Here we are not attempting to overcome such elaborate methods and solutions but intend to show the quantitative drivers of success in soccer from a different angle - as network science sees it. For that, we follow some of the earlier work of Barabasi et. al and others on the quantification of success and the role of networks in team success~\cite{guimera2005team, barabasi2018formula, janosov2020quantifying}.

\paragraph{Player-level success.} To capture the success of players, one could pick a large number of different KPIs depending on the goal they wish to study. Here, to illustrate the role of networks, we picked their current market value as a proxy of their success. We also note that we do not take sides on how much the player's financial success may proxy their talent or their sense of branding, just expect what the market shows.

To assess the players' characteristics, we computed several measures describing them. First, we added measures describing their career: the year of their first transfer, their first public market value, and the number of transfers they had. Then we added several network measures capturing their network positions: Weighted Degree measuring the total weight of connections a player has, Closeness Centrality measuring how few hops a player is from the others, Betweenness Centrality centrality capturing the network bridge positions, and Clustering tells us what fraction of a player's connections know each other.

\begin{figure}[!hbt]
\centering
\includegraphics[width=0.80\textwidth]{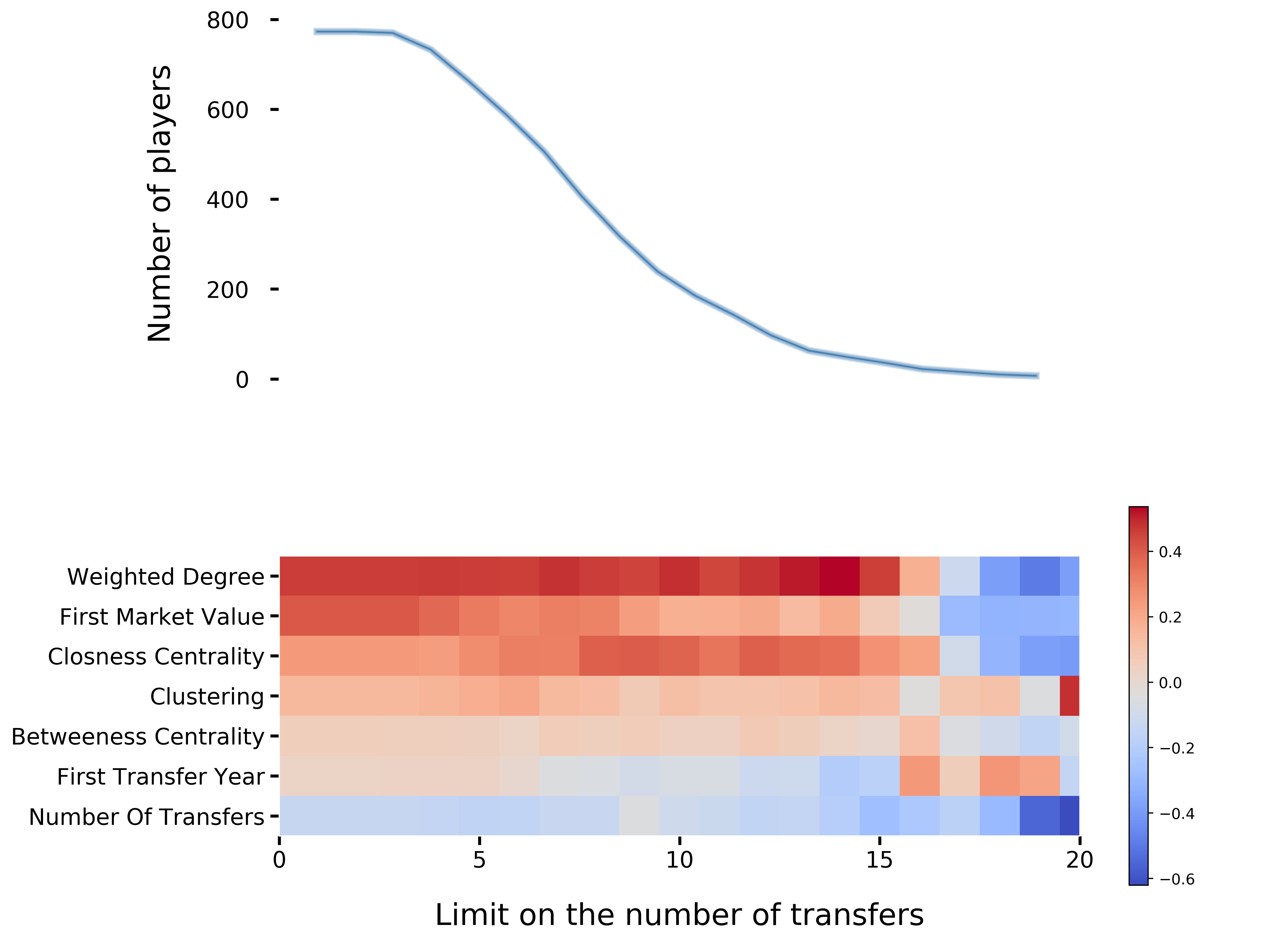}
\caption{The number of players, and the correlations between their market value and the different descriptive features as a function of the number of transfers they had.}
\label{fig:fig3}
\end{figure}

Then, we conducted a series of correlation analyses where we correlated these measures against each player's current market value. Additionally, we differentiated between players based on their seniority, measured by the number of transfers they had. We realize having played for more teams doesn't necessarily mean that a player has had a longer career, younger players can be loaned out to gain experience, but this approach works well with our data set. Figure \ref{fig:fig3} shows the effect of this on the number of players analyzed: less than 1\% of the players have less than 2 transfers, while the line between 7 and 8 transfers splits
the players into two roughly equal parts. Also, slightly less than 5\% of the players have more than 15 transfers which translates into about 30 veterans.

Furthermore, Figure \ref{fig:fig3} also shows the correlations between the players' descriptors against their seniority. We sorted this chart based on the full population of players, meaning that the features in the correlation matrix are put in a descending order based on the correlation matrix when the number of transfers equals zero. The changing patterns in the matrix indicate the different correlation trends as we restrict our analysis to more and more senior players.

The most striking feature of the correlation values, ranging from 0.53 to -0.62, is that no matter the player's seniority, Weighted Degree plays a superior role. One can also notice that the negative correlation values typically occur for the most senior players, and are probably more accounted for noise than signal. For younger players, the graph also shows that after the Weighted Degree comes to their first market value, and then Closeness Centrality. However, at the tenure of about five transfers, the picture changes, and the top three most correlated features to the current market value end up being derived from the network of players. While we are not claiming pinpoint accuracy here, these findings certainly indicate that the role of networks shows significant potential for the expected success of a soccer player.

\paragraph{Country-level success.} Last but not least, we also shoot our shot at understanding some aspects of the final - and certainly not unsurprising - ranking of the World Cup. For this, we used the final ranking of all 32 teams at the 2022 FIFA World Cup~\cite{sportingnews}. Then, we went back to the previously introduced player-level statistics covering the total number of transfers, the first year of transfer, the total market value, and several network measures first attached to each player, and then aggregated to the level of the country's teams. The aggregation, depending on the distribution of the underlying values, happened either by taking the mean or the median values of a country's team members.

Next, we ventured into the world of predicting algorithms. To be more precise, we built a binary XGBoost~\cite{chen2016xgboost} classifier aimed to distinguish between those teams which made it to the Rd of 16 versus those that didn't make it longer than the Groups. We note that even in this scenario, the number of data points was very low. In future research, this part of the analysis could be improved by adding other World Cups to the data, which may also allow one to do more elaborate predictions, covering the Quarterfinals or even the grand finale.

\begin{figure}[!hbt]
\centering
\includegraphics[width=0.7\textwidth]{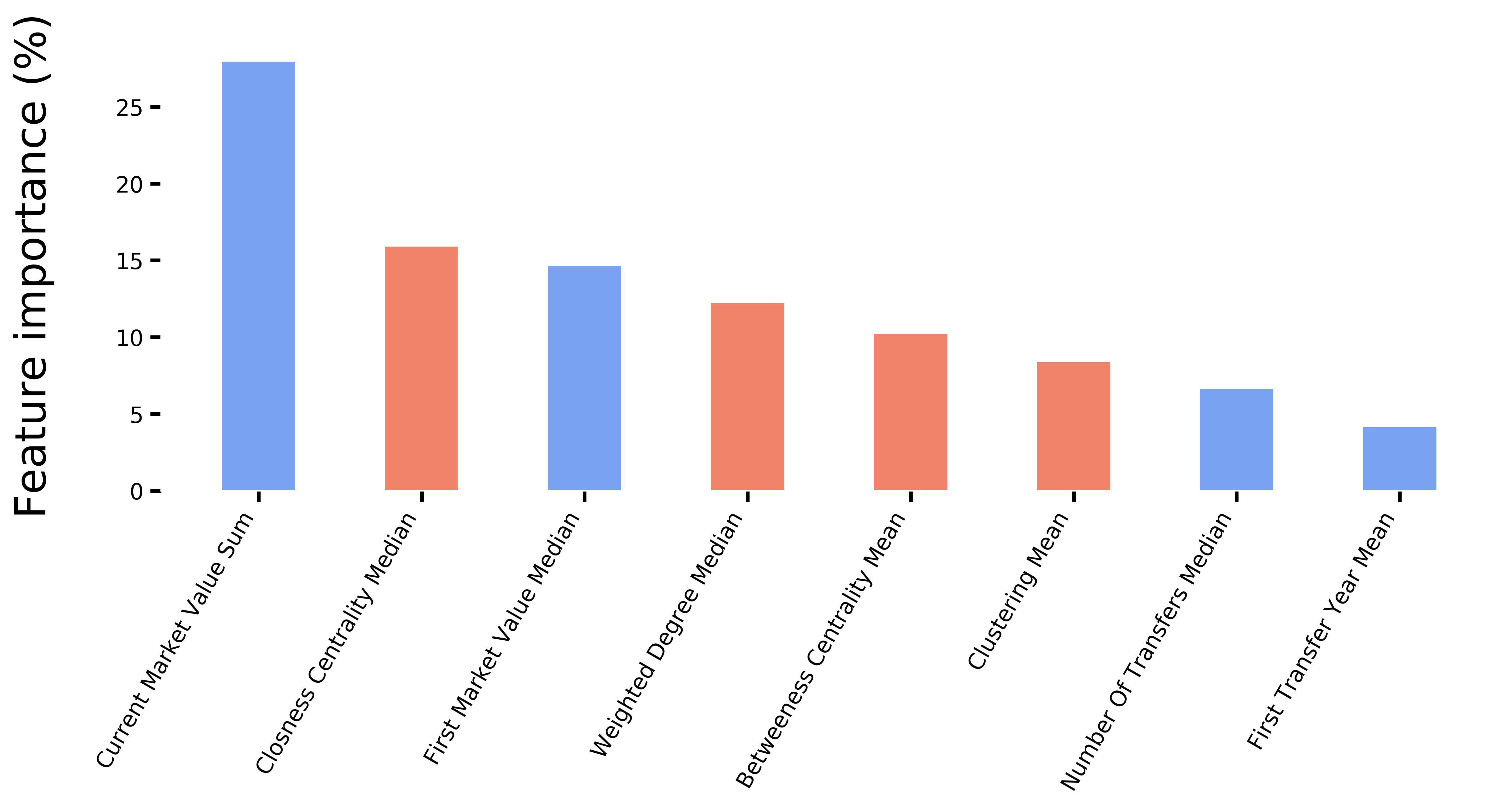}
\caption{Feature importance expressed by relative values when differentiating the teams tho dropped out after the first round from the rest by using binary classification.}
\label{fig:fig4}
\end{figure}

We optimized our XGBoost model using 250 estimators, a maximum tree depth of 6, a learning rate varying between 0.1 and 0.005, a grid search algorithm, and 5-fold cross-validation. On the one hand, the overall prediction accuracy resulted in a fairly modest value of 60\%. Still, it is worth taking a look at the feature importance analysis shown in Figure \ref{fig:fig4}. This figure shows that - probably not so - surprising, the most important driver of a team's position is the total current market value of its players. High paychecks seem to pay off. At least to some extent pays - the correlation between team rank and the total value is still just somewhere between 0.5 and 0.6. This is then followed, somewhat in a tie, between closeness centrality and first market value. While first market value - for more junior players - is certainly correlated with their current value, a network centrality reaching that level certainly emphasizes the role of the network. While the other network measures are performing around 10\% each in terms of their relative importance, we leave it for previous research to nail down the exact meaning of these figures.

\begin{figure}[!hbt]
\centering
\includegraphics[width=0.5\textwidth]{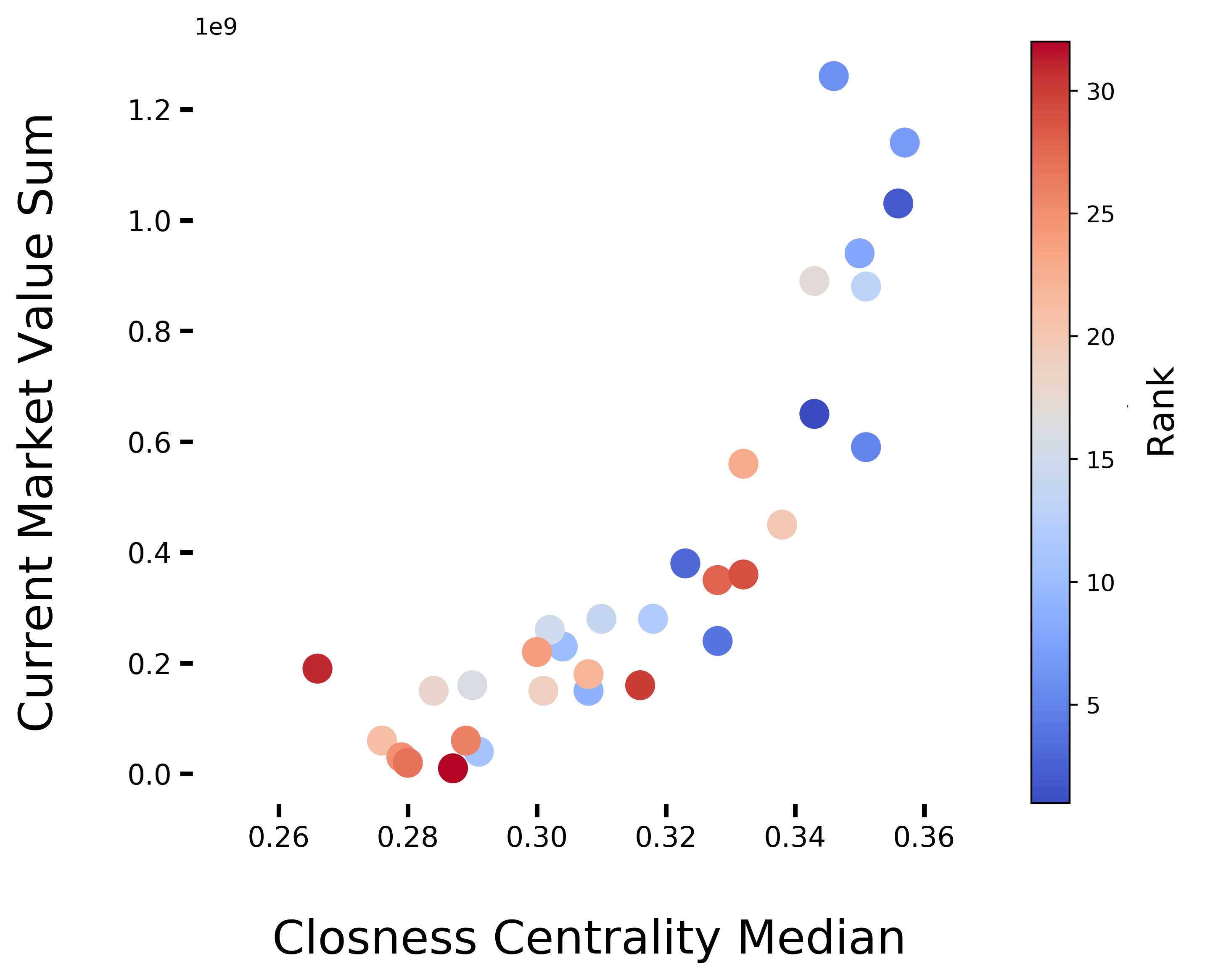}
\caption{On this scatter plot, each team is represented by a dot colored according to its final rank, while its position is determined by the median closeness centrality and the total current market value of its players.}
\label{fig:fig5}
\end{figure}

Finally, to outline the signal behind the noise caused by the low number of data points, we created Figure \ref{fig:fig5} showing how closeness centrality and market value guide the teams in terms of their final rankings.

\section{Conclusion}

In conclusion, we saw in our analysis how network science and visualization can uncover and quantify things that experts may have a gut feeling about but lack the hard data. This depth of understanding of internal and team dynamics that is possible through network science can also be critical in designing successful and stable teams and partnerships. Moreover, this understanding can lead to exact applicable insights on transfer and drafting strategies or even spotting and predicting top talent at an early stage. While this example is about soccer, you could very much adapt these methods and principles to other collaborative domains that require complex teamwork and problem solving with well-defined goals, from creative production to IT product management.


\end{document}